\documentclass[]{article}
\usepackage{times}
\usepackage{cite}
\usepackage{amsmath}
\usepackage{graphicx}
\usepackage{latexsym}

\usepackage{draftwatermark}
\SetWatermarkText{ White-paper}
\begin{document}
\title{Mesh Model (MeMo): A Systematic Approach to Agile System Engineering}
\author{
{\textbf{Amit Kumar Mishra, and Alan Langman}} \\
University of Cape Town\\
South Africa 7700 \\
akmishra@ieee.org
}
\date{}
\maketitle


\section{Introduction}
Innovation and entrepreneurship  have a very special role to play in creating sustainable development in the world. 
Engineering design plays a major role in innovation. 
These are not new facts. However this added to the fact that in  current time knowledge seem to increase at an exponential rate, growing twice every few months \footnote{$bostoncommons.net/knowledge-doubling$}. 

This creates a need to have newer methods to innovate with very little scope to fall short of the expectations from customers. In terms of reliable designing, system design tools and methodologies have been very helpful and have been in use in most engineering industries for decades now. But traditional system design is rigorous and rigid.  

As we can see, we need an innovation system that should be rigorous and flexible at the same time. We take our inspiration from biosphere, where some of the most rugged yet flexible plants are creepers which grow to create mesh. 
In this thematic paper we shall explain our approach to system engineering which we call the MeMo (Mesh Model) that fuses the rigor of system engineering with the flexibility of agile methods to create a scheme that can give rise to reliable innovation in the high risk market of today.

\section{Recent Challenges and  Trends}
The current generation of system engineering (SE) practices have been extremely helpful in managing big, complex and system-of-system natured projects. 
However, as we enter into the exponential zone of knowledge growth the current practices we will have some major limitations. Some of these (taken from INCOSE vision-2025 document) are as follows. 
\begin{itemize}
\item Because of rapid development at component level the old fashioned architecture based system design is not reliable. This means we depend on a system design which emerges from pieces. 
This makes the system difficult to design and verify. 
\item Again, due to the rapid development in different components there is a need for agile change of design architecture. 
However, traditional system engineering practices are not able to handle this agility and thereby causing loss in investment and knowledge. 
\item A holistic approach to SE is good to take care of programmatic side of projects but not the technical side. 
This fosters risky decisions. 
\end{itemize}
Hence, we need an approach which can 
\begin{itemize}
\item be inclusive of rapid technical development; 
\item fosters agility; 
\item has scope to use the knowledge created at project life cycle phase boundaries; and 
\item gives an avenue to reduce risky-decisions by creating enough local feedbacks in the process. 
\end{itemize}
We shall keep the above points in mind while discussing our proposed model. 

It can be noted that there are two major challenges driving recent trends in innovation (e.g. Design Thinking, Innovators Methods, Agile Development etc.). 
First of all is the exponential growth of knowledge and, hence, the coming of newer technologies at a faster pace every year. 
Second is the emergence of demand uncertainty. 
Conventional system engineering and design processes are well adapted to deal with technical uncertainty (in the presence of demand certainty). 
However they are not good enough to handle either demand uncertainty with technological certainty or (even worse) demand undertaking with technological uncertainty.

\section{Spiral meets V to Create a Mesh}
As discussed in the previous section we need a process that is agile enough to run through extreme technical and demand uncertainties. 
One of the important ways to achieve this is to have rapid customer feedback based design thinking. 
This helps us to take care of the demand uncertainty. 
A spiral approach, almost, takes care of technological uncertainties. 
One of the lacuna of the Spiral Model is the lack of a designated verification phase. 

In our proposed model we combine the spiral model with well defined Design Blocks (DB). 
The Design Blocks include  all or some of the design steps like requirement analysis, solution choice, STTPLE analysis, rapid prototype, detailed design and acceptance test procedure definition. 
The choice to run the full Design Block or part of it is up to the technical uncertainty that faces the development. 

In addition to a spiral way of developing the Design Blocks we introduce the Feedback Collection Block (FCB) to take care of the demand uncertainty. 
Each industry can set its own (semi-automated) process to run the FCBs and its own process to store and interpret the results from these. 
It can be noted here that the outcome from FCBs are mostly technology independent. 
Hence, these can be used not only in the immediate-next DB but also by further DBs later in the product life-cycle. 
This also helps in preserving innovations and knowledge across design cycles and versions.

\begin{figure}[th]
\centering
\includegraphics[width=0.8\linewidth]{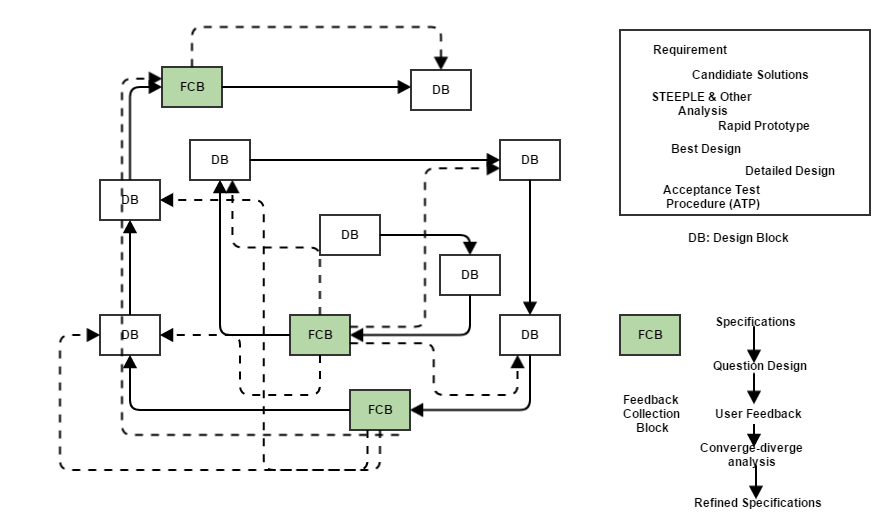}
\caption{Schematic of the ``mesh model''. An important block introduced is the FCB block that helps to design, collect and archive feedback on a regular basis. 
These feedbacks can be used in any of the Design Blocks that come later. This makes sure knowledge and innovation transfer is maintained across different design phases and cycles. 
This also makes sure that the specifications are agile enough to accommodate technological changes that may come over the course of the design-life-cycle of the project. 
Not shown is the fact that these FCB and their outcome can also be stored to help with other parallel projects running in bigger firms.}
\label{mesh}
\end{figure}

\section{Conclusion}
In this brief white paper we have discussed the status quo of engineering innovation and discussed some of the current and upcoming challenges. 
Then we proposed a process for engineering system design and development in the face of demand and technological uncertainty. 
This also takes into account all the key challenges faced by the current system design processes (as detailed in the INCOSE vision document).

\bibliographystyle{IEEEtran}
\bibliography{ref}

%
%

\end{document}